\documentstyle[12pt]{article}
\newcommand{\ba}[1]{\begin{eqnarray} \label{#1}}
\newcommand{\ea}{\end{eqnarray}}
\newcommand{\nn}{\nonumber}
\newcommand{\rf}[1]{(\ref{#1})}
\def\half{{1\over 2}}
\def\CN{{\cal N}}
\def\sb{\sin\!\beta}
\def\cb{\cos\!\beta}
\def\tb{\tan\!\beta}

\def\tw{\tan\!\theta^{}_W}
\begin{document} 

\begin{center}
{\Large \bf About direct Dark Matter detection \\
           in Next-to-Minimal Supersymmetric \\
            Standard Model
\par }

\bigskip

{\large V.A.~Bednyakov\footnote{
        Laboratory of Nuclear Problems,
	Joint Institute for Nuclear Research,
        Moscow region, 141980 Dubna, Russia
	}
	 and H.V.~Klapdor-Kleingrothaus}
\bigskip

        {\it
        Max-Planck-Institut f\"{u}r Kernphysik, \\
        Postfach 103980, D-69029, Heidelberg, Germany
        }

\end{center}

\begin{abstract}
        Direct dark matter detection is considered in the 
	Next-to-Minimal Supersymmetric Standard Model (NMSSM).
	The effective neutralino-quark Lagrangian 
        is obtained and event rates are calculated for the 
	$^{73}$Ge isotope.
	Accelerator and cosmological constraints on the NMSSM
	parameter space are included.
	By means of scanning the parameter space
	at the Fermi scale we show that the lightest neutralino
	could be detected in dark matter 
	experiments with sizable event rate.

\noindent
PACS number(s): 14.80.Ly, 95.35+d, 98.80Cq
\end{abstract}

\bigskip

\section{Introduction}
        In not too far future new very
        sensitive dark matter (DM) detectors
\cite{GENIUS}--\cite{Baudis}
	may start to operate,
	and 
        one expects new, very important data from these experiments.
        The future experimental progress forces investigators to
        know better the variety and property of the
        dark matter particles.
        The lightest supersymmetric particle (LSP), the neutralino,
        is considered now as a most promising candidate, 
	which may compose the main fraction of so-called cold dark matter.
        The prospects of the direct and indirect detections
        of the LSP have comprehensively been investigated
\cite{JungKamGriest} 
        in the various versions of the Minimal Supersymmetric
        Standard Model (MSSM) 
\cite{HaberKane}.

        In this paper we consider direct detection of this relic LSP in 
        the Next-To-Minimal Supersymmetric Standard Model (NMSSM) 
\cite{Drees, EGHRZ}.
	The Higgs sector of the NMSSM contains
	five physical neutral Higgs bosons, three Higgs scalars,
	two pseudoscalars, and two
	degenerate physical charged Higgs particles $C^\pm$.
	The neutralino sector is extended to five neutralinos instead
        of four in the MSSM.
	The remaining particle content is identical with that of the MSSM.

        The NMSSM is mainly motivated by its potential to eliminate the
	so-called $\mu$ problem of the MSSM 
\cite{KimNillis}, where the origin of the 
	$\mu$ parameter in the superpotential
$ W_{\mbox{\scriptsize MSSM}}=\mu  H_1 H_2 $ 
	is not understood.
	For phenomenological reasons it has to be of the order
	of the electroweak scale, while the "natural" mass scale would be
	of the order of the GUT or Planck scale.
	This problem is evaded in the NMSSM where the
	$\mu$ term in the superpotential is dynamically generated through
	$\mu = \lambda x$ with a dimensionless coupling $\lambda$ and
	the vacuum expectation value $x$ of the Higgs singlet.
        Another essential feature of the NMSSM is the fact that the 
	mass bounds for the Higgs bosons and neutralinos are weakened.
	While in the MSSM experimental data imply a lower mass bound
        of about 20 GeV for the LSP 
\cite{ALEPH}, 
	very light or massless neutralinos and Higgs bosons
	are not excluded in the NMSSM 
\cite{FrankeFraasBartl,FrankeFraas1}.
	Furthermore the upper tree level mass bound for the
	lightest Higgs scalar of the MSSM
\begin{equation} \label{tlb}
	m_h^2 \leq m_Z^2 \cos^22\beta
\end{equation}
	is increased to
$ m_{S_1}^2 \le m_Z^2 \cos^2 2\beta+\lambda^2(v_1^2+v_2^2) \sin^2 2\beta$.
	Taking into account the weak coupling
	of the singlet Higgs the NMSSM may still remain a
        viable model when the MSSM can be ruled out due to 
(\ref{tlb}).

	The above arguments make an intensive study of the
	NMSSM phenomenology very desirable.
	Previously the Higgs and neutralino sectors
	of the NMSSM were carefully studied in
\cite{FrankeFraasBartl}%
--\cite{Pandita}.
	The calculation of the LSP relic abundance 
	in the NMSSM was performed for the first time in
\cite{ApelSW} and recently in  
\cite{AStephan}.
\smallskip

	The outline of this paper is as follows.
	In 
Sec.~\ref{sec:Lag} we describe the Lagrangian of the NMSSM.
	Since the additional singlet superfield of the NMSSM leads
	to extended Higgs and neutralino sectors,
	we present the Higgs and neutralino mixings.
Section~\ref{sec:dark} collects formulas relevant for 
	calculation of the event rate for direct dark matter 
	detection in the framework of the NMSSM.
	In 
Sec.~\ref{sec:constraints} we discuss the constraints on 
	the NMSSM parameter space which are used in our analysis.
	In 
Sec.~\ref{sec:numerical} we shortly describe our numerical 
	procedure and discuss the results obtained.
Sec.~\ref{sec:conclusion} contains a conclusion.

\section{The Lagrangian of the NMSSM} \label{sec:Lag}
        The NMSSM superpotential is 
\cite{FrankeFraas} 
	($\varepsilon_{12}=-\varepsilon_{21}=1$):
\begin{eqnarray}
W & = & \lambda \varepsilon_{ij} H_1^i H_2^j N -\frac{1}{3} k N^3
\nonumber \\ & &
+h_u \varepsilon_{ij} \tilde{Q}^i \tilde{U} H_2^j
-h_d \varepsilon_{ij} \tilde{Q}^i \tilde{D} H_1^j
-h_e \varepsilon_{ij} \tilde{L}^i \tilde{R} H_1^j,
\end{eqnarray}
        where $H_1 = (H_1^0,H^-)$ and
	      $H_2=(H^+,H_2^0)$ are the SU(2) Higgs
	doublets with hypercharge $-1/2$ and $1/2$ and
	$N$ is the Higgs singlet with hypercharge 0.
	The notation of the fermion doublets and singlets
	is conventional, generation indices are omitted. 
	Contrary to the MSSM, the superpotential of the NMSSM
	consists only of trilinear terms with dimensionless couplings.

	The electroweak gauge-symmetry $SU(2)_I \times U(1)_Y$
	is spontaneously broken to the electromagnetic
	gauge-symmetry $U(1)_{em}$ by the Higgs VEVs
	$\langle H_i^0 \rangle = v_i$  with $i = 1,2$
	and $\langle N \rangle = x$,
	where $v = \sqrt{v_1^2 + v_2^2} = 174$~GeV,
	$\tan \beta = v_2/v_1$.

        The most general supersymmetry breaking potential
	can be written as
\cite{FrankeFraas}
\begin{eqnarray}
\label{vs}
V_{\mbox{soft}} & = &
m_1^2 |H_1|^2 + m_2^2 |H_2|^2+m_3^2 |N|^2
\nonumber \\ & &
+m_Q^2 |\tilde{Q}|^2 + m_U^2 |\tilde{U}|^2 + m_D^2 |
\tilde{D}|^2
+m_L^2 |\tilde{L}|^2 + m_E^2 |\tilde{R}|^2
\nonumber \\  & &
-(\lambda A_\lambda \varepsilon_{ij} H_1^i H_2^j N + \mbox{h.c.})
-(\frac{1}{3}kA_k N^3 + \mbox{h.c.})
\nonumber \\  & &
+(h_u A_U \varepsilon_{ij} \tilde{Q}^i \tilde{U} H_2^j
-h_d A_D \varepsilon_{ij} \tilde{Q}^i \tilde{D} H_1^j
-h_e A_E \varepsilon_{ij} \tilde{L}^i \tilde{R} H_1^j
+\mbox{h.c.})
\nonumber \\ & &
+\frac{1}{2}M \lambda^a \lambda^a
+\frac{1}{2}M' \lambda ' \lambda '.
\end{eqnarray}

        As free parameters appear
	the ratio of the doublet vacuum expectation values,
	$\tan\beta$, the singlet vacuum expectation value $x$,
	the couplings in the superpotential $\lambda$ and $k$,
	the parameters $A_\lambda$, $A_k$,
	as well as $A_U$, $A_D$, $A_E$ (for three generations)
	in the supersymmetry breaking potential, the gaugino mass
	parameters $M$ and $M'$, and the scalar mass parameters
	for the Higgs bosons $m_{1,2,3}$, squarks $m_{Q,U,D}$ and sleptons
	$m_{L,E}$.

\smallskip
	The minimization conditions for the scalar potential
$\partial V / \partial v_{1,2} =0$, $\partial V / \partial x =0$
	eliminate three parameters of the Higgs sector
	which are normally chosen to be $m_1^2$, $m_2^2$ and $m_3^2$.
	Then at tree level the elements of the symmetric CP-even mass
	squared matrix ${\cal M}_S^2=({M_{ij}^S}^2)$ become in the basis
$(H_1,H_2,N)$
\begin{eqnarray*}
{M_{11}^S}^2 & = & \frac{1}{2} v_1^2 ({g'}^2+g^2) +
\lambda x\tan\beta (A_\lambda + kx), \\
{M_{12}^S}^2 & = & -\lambda x (A_\lambda + kx)
+v_1v_2(2\lambda^2-\frac{1}{2} {g'}^2 - \frac{1}{2} g^2 )\\
{M_{13}^S}^2 & = & 2 \lambda ^2 v_1 x -2\lambda kxv_2 -
\lambda A_\lambda v_2, \\
{M_{22}^S}^2 & = & \frac{1}{2} v_2^2 ({g'}^2+g^2) +
\lambda x\cot\beta (A_\lambda + kx), \\
{M_{23}^S}^2 & = & 2 \lambda ^2 v_2 x -2\lambda kxv_1 -
\lambda A_\lambda v_1, \\
{M_{33}^S}^2 & = & 4k^2x^2-kA_kx+\frac{\lambda A_\lambda v_1 v_2}{x}.
\end{eqnarray*}
	In the same way one finds for the elements of
	the CP-odd matrix ${\cal M}_P^2$
\begin{eqnarray*}
&{M_{11}^P}^2  =  \lambda x(A_\lambda + kx) \tan\beta , 
&{M_{12}^P}^2  =  \lambda x(A_\lambda + kx)  ,\\
&{M_{13}^P}^2  =  \lambda v_2 (A_\lambda - 2kx), 
&{M_{22}^P}^2  =  \lambda x(A_\lambda + kx) \cot\beta ,\\
&{M_{23}^P}^2  =  \lambda v_1 (A_\lambda - 2kx), 
&{M_{33}^P}^2  =  \lambda A_\lambda \frac{v_1v_2}{x} + 4\lambda k v_1 v_2
+3kA_kx ,
\end{eqnarray*}
	and for the charged Higgs matrix one obtains
$$
{\cal M}_c^2 = \left( \lambda A_\lambda x + \lambda k x^2 - v_1 v_2
\left( \lambda^2 -\frac{g^2}{2} \right) \right)
\left( \begin{array}{cc} \tan\beta & 1 \\ 1 & \cot\beta
\end{array} \right).
$$
	In our numerical analysis 
	we have included 1-loop radiative corrections
	to Higgs mass matrices following 
\cite{ElliottKingWhite,KingWhite-rc}.

	Assuming CP conservation in the Higgs sector,
	the Higgs matrices are diagonalized by the real
	orthogonal $3\times 3$ matrices $U^S$ and $U^P$, respectively,
\begin{eqnarray*}
\mbox{Diag}(m_{S_1}^2,m_{S_2}^2,m_{S_3}^2) & = & {U^S}^T {\cal M}_S^2
U^S, \\
\mbox{Diag}(m_{P_1}^2,m_{P_2}^2,0) & = & {U^P}^T {\cal M}_P^2 U^P,
\end{eqnarray*}
	where $m_{S_1} < m_{S_2} < m_{S_3}$ and
	$m_{P_1}<m_{P_2}$ denote the masses of  
	the mass eigenstates of the neutral scalar 
	Higgs bosons $S_a$ $(a=1,2,3)$ 
	and neutral pseudoscalar Higgs bosons $P_\alpha$ $(\alpha = 1,2)$
\cite{FrankeFraas}.

	With fixed parameters of the Higgs sector
	the masses and mixings of the neutralinos are determined by the two
	further parameters $M$ and $M'$ of the Lagrangian
\begin{eqnarray*}
   {\cal L} = -\frac{1}{2} \Psi^T M \Psi + {\rm h.c.},
	\qquad
   \Psi^T = (-i \lambda_1,-i \lambda^3_2,
   \Psi^0_{H_1},\Psi^0_{H_2},\Psi_N).
\end{eqnarray*}
\noindent
	In this basis the symmetric mass matrix $M$ of
	the neutralinos has the form:
{\small
\[ \left( \begin{array}{ccccc}
   M^\prime  &  0  &  - m_Z \: \sin \theta_W \: \cos \beta  &
   m_Z \: \sin \theta_W \: \sin \beta  &  0  \\

   0  &  M  &  m_Z \: \cos \theta_W \: \cos \beta  &
   - m_Z \: \cos \theta_W \: \sin \beta  &  0  \\

   - m_Z \: \sin \theta_W \: \cos \beta  &
   m_Z \: \cos \theta_W \: \cos \beta  &  0  &
   \lambda \: x  &  \lambda \: v_2  \\

   m_Z \: \sin \theta_W \: \sin \beta  &
   - m_Z \: \cos \theta_W \: \sin \beta  &
   \lambda \: x  &  0  &  \lambda \: v_1  \\

   0  &  0  &  \lambda \: v_2  &  \lambda \: v_1  &  - 2 \: k \: x

   \end{array}\right). \]
\par}

	The mass of the neutralinos is obtained by diagonalizing
	the mass matrix $M$ with the orthogonal matrix $N$:
\begin{eqnarray*}
   {\cal L} = -\frac{1}{2} \: m_i \:
   \overline{\tilde{\chi}^0_i} \: \tilde{\chi}^0_i,
\qquad
   \tilde{\chi}^0_i = \left(\begin{array}{cc}
   \chi^0_i  \\ \overline{\chi}^0_i
   \end{array}\right) \\
\nonumber
\mbox{with} \:\:
   \chi^0_i = {\cal N}_{ij} \Psi_j
	\:\:\: {\rm and} \:\:\:
   M_{\rm diag} = N \: M \: N^{T}.
\end{eqnarray*}
\noindent
	The neutralinos $\tilde{\chi}^0_i$ ($i$ = 1--5) are ordered
	with increasing mass $|m_i|$, thus $\chi\equiv
	\tilde{\chi}^0_1$ is the LSP neutralino.
	The matrix elements ${\cal N}_{i j}$ ($i,j$ = 1--5)
	describe the composition of the neutralino $\tilde{\chi}^0_i$ 
	in the basis $\Psi_j$.
	For example the bino fraction of the lightest neutralino is given
	by ${\cal N}^2_{1 1}$ and the singlino fraction of this neutralino
	by ${\cal N}^2_{1 5}$.

\section{Neutralino-nucleus elastic scattering} \label{sec:dark}
        A dark matter event is elastic scattering of a DM neutralino from
   	a target nucleus producing a nuclear recoil which can be detected
	by a suitable detector.
        The corresponding event rate depends on the distribution of
        the DM neutralinos in the solar vicinity and
        the cross section of  neutralino-nucleus elastic scattering.

        The relevant low-energy effective neutralino-quark 
	Lagrangian can be written in a general form as
\cite{JungKamGriest,DreesNojiriER,EllisFlores,NathArnowitt}
\begin{equation} \label{Lagr}
  L_{eff} = \sum_{q}^{}\left( {\cal A}_{q}\cdot
      \bar\chi\gamma_\mu\gamma_5\chi\cdot
                \bar q\gamma^\mu\gamma_5 q +
    \frac{m_q}{M_{W}} \cdot{\cal C}_{q}\cdot\bar\chi\chi\cdot\bar q q\right)
      \ +\ O\left(\frac{1}{m_{\tilde q}^4}\right),
\end{equation}
        where terms with vector and pseudoscalar quark currents are
        omitted being negligible in the case of non-relativistic
        DM neutralinos with typical velocities $v \approx 10^{-3} c$.

	The coefficients in the effective Lagrangian
\rf{Lagr} have the form:
\ba{Aq1}
 {\cal A}_{q} =
	&-&\frac{g_{2}^{2}}{4M_{W}^{2}}
	   \Bigl[\frac{{\cal N}_{14}^2-{\cal N}_{13}^2}{2}T_3 \nn \\
        &-& \frac{M_{W}^2}{m^{2}_{\tilde{q}1} - (m_\chi + m_q)^2}
	   (\cos^{2}\theta_{q}\ \phi_{qL}^2
	   + \sin^{2}\theta_{q}\ \phi_{qR}^2) \nn \\
        &-& \frac{M_{W}^2}{m^{2}_{\tilde{q}2} - (m_\chi + m_q)^2}
	     (\sin^{2}\theta_{q}\ \phi_{qL}^2
	     + \cos^{2}\theta_{q}\ \phi_{qR}^2) \nn \\
        &-& \frac{m_{q}^{2}}{4}P_{q}^{2}\left(\frac{1}{m^{2}_{\tilde{q}1}
		- (m_\chi + m_q)^2}
             + \frac{1}{m^{2}_{\tilde{q}2}
                - (m_\chi + m_q)^2}\right) \nn \\
        &-& \frac{m_{q}}{2}\  M_{W}\  P_{q}\  \sin2\theta_{q}\
            T_3 ({\cal N}_{12} - \tan\theta_W {\cal N}_{11}) \nn \\
	&\times&\left( \frac{1}{m^{2}_{\tilde{q}1}- (m_\chi + m_q)^2}
    - \frac{1}{m^{2}_{\tilde{q}2} - (m_\chi + m_q)^2}\right)\Bigr]
\ea
\ba{Cq1}
 {\cal C}_{q} =
	&-&  \frac{g_2^2}{4} \Bigl[
         - \sum^{}_{a=1,2,3} Q^{L\prime\prime}_{a11}
                 \frac{1}{m^2_a} {\cal V}_{aq}
\nn \\
	&+& P_q \left(\frac{\cos^{2}\theta_{q}\ \phi_{qL} -
     \sin^{2}\theta_{q}\ \phi_{qR}}{m^{2}_{\tilde{q}1} - (m_\chi + m_q)^2}
         -\frac{\cos^{2}\theta_{q}\ \phi_{qR} -
     \sin^{2}\theta_{q}\ \phi_{qL}}{m^{2}_{\tilde{q}2} - (m_\chi +
	m_q)^2}\right) \nn \\
	&+& \sin2\theta_{q}(\frac{m_q}{4 M_W} P_{q}^{2} -
                           \frac{M_W}{m_q} \phi_{qL}\ \phi_{qR}) \nn \\
	&\times&\left(\frac{1}{m^{2}_{\tilde{q}1} - (m_\chi + m_q)^2} -
\frac{1}{m^{2}_{\tilde{q}2} - (m_\chi + m_q)^2}\right)\Bigr].
\ea
     Here
\begin{eqnarray*}
{\cal V}_{aq}
        &=&  
           \Bigl[
                (\half+T^{}_{3q}) \frac{U^S_{a2}}{\sb}
               +(\half-T^{}_{3q}) \frac{U^S_{a1}}{\cb}
           \Bigr],
\\
  Q^{L\prime\prime}_{a11}
        &=&  
	(\CN_{12} - \tw \CN_{11})
            \bigl[
                  U^S_{a1} \CN_{13} - U^S_{a2} \CN_{14}
            \bigr]
\\
       &+&  \sqrt{2} \lambda\CN_{15}
            \bigl[
                  U^S_{a1} \CN_{14} + U^S_{a2} \CN_{13}
            \bigr]
	    - 2\sqrt{2} k U^S_{a3}\CN_{15}^2,
\\
\phi_{qL} &=& {\cal N}_{12} T_3 + {\cal N}_{11}(Q -T_3)\tan\theta_{W},
\\
\phi_{qR} &=& \tan\theta_{W}\  Q\  {\cal N}_{11},
\\
P_{q} &=&  \bigl(\frac{1}{2}+T_3\bigr) \frac{{\cal N}_{14}}{\sin\beta}
          + \bigl(\frac{1}{2}-T_3\bigr) \frac{{\cal N}_{13}}{\cos\beta}.
\end{eqnarray*}

        The coefficients ${\cal A}_q$ and ${\cal C}_q$
	take into account squark mixing
   	$\tilde{q}_L-\tilde{q}_R$ and the contributions of
        all CP-even Higgs bosons.
	Under the assumption $\lambda=k=0$ these formulas coincide 
	with the relevant formulas in the MSSM 
\cite{DreesNojiriER}.

        A general representation of the differential cross section
        of neutralino-nucleus scattering can be given in terms of
        three spin-dependent ${\cal  F}_{ij}(q^2)$ and
        one spin-independent ${\cal F}_{S}(q^2)$ form factors as follows
\cite{EngelVogel}
\ba{cs}
\frac{d\sigma}{dq^2}(v,q^2)&=&\frac{8 G_F}{v^2} \Bigl(
   		a_0^2\cdot {\cal F}_{00}^2(q^2)
	     + a_0 a_1 \cdot {\cal F}_{10}^2(q^2) \nn \\
	     & &\hphantom{\frac{8 G_F}{v^2}}
   	      + a_1^2\cdot {\cal F}_{11}^2(q^2)
   	      + c_0^2\cdot A^2\ {\cal F}_{S}^2(q^2)
   	\Bigr).
\ea
   	The last term corresponding to the spin-independent scalar
	interaction gains coherent enhancement $A^2$
	($A$ is the atomic weight of the nucleus in the reaction).
   	The coefficients $a_{0,1}, c_0$ do not depend on nuclear structure
   	and relate to the parameters ${\cal A}_q$, ${\cal C}_q$
	of the effective Lagrangian
\rf{Lagr} and to parameters 
   	characterizing the nucleon structure.
	In what follows we use notations and definitions of our paper 
\cite{Superlight}.	
 
        An experimentally observable quantity is the differential 
	event rate per unit mass of the target material
$$
\frac{dR}{dE_r} = {\Bigl[ N \frac{\rho_\chi}{m_\chi} \Bigr]}
    \int^{v_{max}}_{v_{min}} dv f(v) v 
\frac{d\sigma}{dq^2} (v, E_r),
	\qquad q^2 = 2 M_A E_r.
$$
   	Here $f(v)$ is the velocity distribution of neutralinos
   	in the earth's frame which is usually assumed to be a
	Maxwellian distribution in the galactic frame.
	$N$ is the number density of the target nuclei. 
$v_{max} = v_{esc} \approx$ 600 km/s and
$\rho_{\chi}$ = 0.3 GeV$\cdot$cm$^{-3}$  are the escape velocity
	and the mass density of the relic neutralinos in
   	the solar vicinity;
$v_{min} = \left(M_A E_r/2 M_{red}^2\right)^{1/2}$ with
$M_A$ and $M_{red}$ being the mass of nucleus $A$ and the reduced
   	mass of the neutralino-nucleus system, respectively.
	
   	The differential event rate is the most appropriate quantity
	for comparing with the observed recoil spectrum and allows one
	to take properly into account spectral characteristics of a
	specific detector and to separate the background.
   	However, in many cases the total  event rate $R$\
   	integrated over the wholere kinematic domain of the recoil energy
   	is sufficient.
   	It is widely employed in theoretical papers for estimating the
	prospects for DM detection,
   	ignoring experimental complications which may occur on the  way.
        In the present paper we are going  to perform a general
	analysis aimed at searching for domains with 
	large values of the event rate $R$\ like those reported in
\cite{GondoloBerg}.
   	This is the reason why we use in the analysis the total
	event rate $R$.

\section{Constraints on the NMSSM parameter space}
\label{sec:constraints}
  	 Assuming that the neutralinos form a dominant part of
    	 the DM in the universe one obtains a cosmological constraint
   	 on the neutralino relic density.
   	 The present lifetime of the universe is at least $10^{10}$ years,
   	 which implies an upper limit on the expansion rate and
   	 correspondingly on the total relic abundance.
   	 Assuming $h_0>0.4$ one finds that  the contribution of
  	 each relic particle species $\chi$  has to obey
~\cite{kolb}:
$
 \Omega_\chi h^2_0<1,
$
   	where the relic density parameter  $\Omega_\chi = \rho_\chi/\rho_c$
   	is the ratio of the relic neutralino mass  density $\rho_\chi$
   	to  the critical one
   	$\rho_c = 1.88\cdot 10^{-29}$h$^2_0$g$\cdot$cm$^{-3}$.

   	We calculate $\Omega_{\chi} h^2_0$  following the standard
   	procedure on the basis of the approximate formula
\cite{Hagelin,DreesNojiriOM}:
\begin{eqnarray}
\label{omega}
\Omega_{\chi} h^2_0 &=& 2.13\times 10^{-11}
\left(\frac{T_{\chi}}{T_{\gamma}}\right)^3
\left(\frac{T_{\gamma}}{2.7K^o}\right)^3
\times N_F^{1/2}
\left(\frac{{\mbox{GeV}}^{-2}}{a x_F + b x_F^2/2}\right).
\end{eqnarray}
   	Here $T_{\gamma}$ is the present day photon temperature,
   	$T_{\chi}/T_{\gamma}$ is the reheating factor,
   	$x_F = T_F/m_{\chi} \approx 1/20$, $T_F$ is
   	the neutralino freeze-out temperature, and $N_F$ is the total
   	number of degrees of freedom at $T_F$. 
	The coefficients $a, b$ are determined from the
   	non-relativistic expansion
$ \langle \sigma_{ann.} v \rangle  \approx a + b x $
   	of the thermally averaged cross section of 
	neutralino annihilation in the NMSSM.
   	We adopt an approximate treatment not taking into account
   	complications, which occur when the expansion fails
\cite{omega}. 
	We take into account all possible channels of
   	the $\chi$-$\chi$ annihilation. 
	The complete list of the relevant formulas in the NMSSM 
   	can be found in 
\cite{AStephan}.

   	Since the neutralinos are mixtures of gauginos,
   	higgsinos, and singlino 
	the annihilation can occur both, via
   	s-channel exchange of the $Z^0$ and Higgs bosons and
   	t-channel exchange of a scalar particle, like a 
	selectron. 
   	This constrains the parameter space, as discussed by
   	many groups
\cite{DreesNojiriOM,relic,relictst}.

        In the analysis we ignore possible rescaling of the local
        neutralino density $\rho$ which may occur in the region of the
        NMSSM parameter space where $\Omega_\chi h^2_0 < 0.025$
\cite{GelminiGonfoloRoulet,Botino,BBEFMS}.  
        If the neutralino is accepted as a dominant part of the DM its
        density has to exceed the quoted limiting value 0.025.
        Otherwise the  presence of additional DM components should be
        taken into account, for instance, by the mentioned rescaling ansatz.  
	However, the halo density is known to be very uncertain. 
	Therefore, one can expect that the rescaling takes place in a 
	small domain of the parameter space. 
	Another point is that the SUSY solution  of the DM problem
	with such low neutralino density becomes questionable.
        We assume neutralinos to be a dominant component of
        the DM halo of our galaxy with a density
        $\rho_{\chi}$ = 0.3 GeV$\cdot$cm$^{-3}$ in the solar vicinity
        and disregard in the analysis points with
        $\Omega_{\chi}h^2_0 < 0.025$.
\smallskip

	The parameter space of the NMSSM and the masses of the 
	supersymmetric particles are constrained by the results 
	from the high energy colliders
	LEP at CERN and Tevatron at Fermilab
\cite{FrankeFraasBartl,FrankeFraas1}.
	A key role for the production of Higgs bosons at
	$e^+e^-$ colliders plays the Higgs coupling to
	$Z$ bosons, while neutralino production at LEP
	crucially depends on the $Z\tilde{\chi}^0\tilde{\chi}^0$ 
	coupling which is formally identical in NMSSM and MSSM 
	and differs only by the neutralino mixing.
	All those couplings are suppressed in the NMSSM if the respective
	neutralinos or Higgs bosons have significant singlet components.
	Therefore NMSSM neutralino and Higgs mass bounds are much weaker
	than in the minimal model
\cite{FrankeFraas}.
	The consequences from the negative neutralino search at LEP
	for the parameter space and the neutralino masses 
	have been studied in 
\cite{FrankeFraasBartl}.
	In 
\cite{FrankeFraas} it is shown that a very
	light NMSSM neutralino cannot even be ruled out at LEP2.

	We used the following constraints from LEP.
	For new physics contributing to the total $Z$ width
$\Delta \Gamma(Z\to \tilde{\chi}^+\tilde{\chi}^- 
	+ Z\to \tilde{\chi}^0_i\tilde{\chi}^0_j) < 23$~MeV.
	For new physics contributing to the invisible $Z$ width
$\Delta \Gamma(Z\to \tilde{\chi}^0_i\tilde{\chi}^0_j) < 8$~MeV.
	From the direct neutralino search
$B (Z \rightarrow \tilde{\chi}^0_1 \tilde{\chi}^0_j) < 2 \times 10 ^{-5}$
	for $j=2,\ldots,5$, 
	and 
$B (Z \rightarrow \tilde{\chi}^0_i \tilde{\chi}^0_j)
	< 5 \times 10 ^{-5}$, for $i,j=2,\ldots,5$.
	The results of LEP searches for $S^{}_a Z$ and $S^{}_a Z^*$ 
	productions 
\cite{ALEPH}, which impose restrictions on the $S^{}_a ZZ$ couplings 
	were included in our analysis.  
	We have also included 
	the experimental bounds from the direct search for
	pseudoscalar Higgs bosons produced 
	together with a Higgs scalar at LEP
\cite{ALEPH}, 
	but this in accordance with 
\cite{FrankeFraas}
	does not significantly affect the excluded parameter domain.

	In our numerical analysis we use the following 
	experimental restrictions for the SUSY particle spectrum in the
	NMSSM:
$ m_{\tilde{\chi}^+_1} \geq 90$~GeV for charginos,
$ m_{\tilde{\nu}}      \geq 80$~GeV for sneutrinos,
$ m_{\tilde{e}_R}      \geq 80$~GeV for selectrons,
$ m_{\tilde{q}}       \geq 150$~GeV  for squarks,
$ m_{\tilde{t}_1}      \geq 60$~GeV  for light stop,   
$ m_{H^+}              \geq 65$~GeV  for charged Higgses and 
$ m_{S_1}              \geq 1$~GeV 
	for the light scalar neutral Higgs.
	In fact, it appeared that all above-mentioned 
	constraints do not allow 
	$m_{S_1}$ to be smaller then 20 GeV.

\section{Numerical Analysis} \label{sec:numerical}

	Randomly scanned parameters of the NMSSM 
	at the Fermi scale are the following:
	the gaugino mass parameters $M^\prime$ and $M$, 
	the ratio of the doublet vacuum expectation values, $\tan\beta$, 
	the singlet vacuum expectation value $x$,
	the couplings in the superpotential $\lambda$ and $k$, 
	squared squark mass parameters $m^2_{Q_{1,2}}$ for the 
	first two generations and $m^2_{Q_3}$ for the third one,
	the parameters $A_\lambda$, $A_k$, 
	as well as $A_t$ for the third generation.
	The parameters are varied in the intervals given below
\begin{center}
\begin{tabular}{c@{~~$<$~~}c@{~~$<$~~}c}
  $-1000$~GeV  & $M^\prime$ &    1000~GeV   \\
  $-2000$~GeV  & $M$        &    2000~GeV   \\
        1      & $\tb$      &      50   \\
        0~GeV  & $x$        &   10000~GeV   \\
   $-0.87$     & $\lambda$  &    0.87   \\
   $-0.63$     & $k$        &    0.63   \\
  100~GeV$^2$  & $m^2_{Q_{1,2}} $    & 1000000~GeV$^2$  \\
  100~GeV$^2$  & $m^2_{Q_3} $    & 1000000~GeV$^2$  \\
  $-2000$~GeV  & $A_t$       &    2000~GeV   \\
  $-2000$~GeV  & $A_\lambda$ &    2000~GeV   \\
  $-2000$~GeV  & $A_k$       &    2000~GeV.   \\
\end{tabular}
\end{center}
	For simplicity the other sfermion mass parameters 
	$m^2_{U_{1,2}}$, $m^2_{D_{1,2}}$, 
	$m^2_{L_{1,2}}$, $m^2_{E_{1,2}}$, and $m^2_{T}$, $m^2_{B}$, 
	$m^2_{L_{3}}$, $m^2_{E_{3}}$, are chosen to be 
	equal to $m^2_{Q_2}$ and $m^2_{Q_3}$, respectively.
	Therefore masses of the sfermions in the same generation 
	differ only due to the D-term contribution.
	Other parameters (except $A_t$) of the supersymmetry 
	breaking potential $A_U$, $A_D$, $A_E$ 
	(for all three generations) are fixed to be zero.

\smallskip

   	The main results of our scan are presented in
Fig.~1  in the form of scatter plots.
	Given in Fig.~1 
	are the total event rate $R$ for
	$^{73}$Ge, and the LSP gaugino 
	fraction ($\CN^2_{11}+\CN^2_{12}$), singlino 
	fraction ($\CN^2_{15}$), 
	and finally relic density parameter
	$\Omega_{\chi}h^2_0$
	versus the LSP mass.
	The left panel in Fig.~1 presents the above-mentioned
	observables obtained without taking into account
	the cosmological relic density constraint.

	In this case the total expected event rate $R$ reaches
	values up to about 50 events per day and per 
	1 kg of the $^{73}$Ge isotope. 
	As on can see from Fig.~1 
	the small-mass LSP (less then about 100~GeV) are mostly
	gauginos, with very small admixture of the singlino component.
	Large masses of the LSP (larger then 100~GeV)
	correspond to sizable gaugino and singlino components
	together perhaps with some higgsino fraction.

	The results of implementation of the
	cosmological constraint 
$$
0.025 < \Omega_\chi h^2_0<1.
$$
	one can see in the right panel of Fig.~1.
	There is approximately a 5-fold reduction of the
	number of the points which fulfill all restrictions in this case.
	Nevertheless quite large values of event rate $R$
	(above 1 event/day/kg) still survive the cosmological
	constraint.
	The lower bound for the mass of the LSP now becomes
	about 3-5~GeV.
	The gaugino component becomes more significant, but
	the singlino fraction can not be completely ruled out
	especially for large masses of the LSP.
	The higgsino component of the LSP remains still possible
	only for LSP masses in the vicinity of $M_Z$.	

	For illustration in 
Fig.~2  we present the calculated event rate $R$\ as function of
	the mass of the lightest scalar Higgs boson, $m^{}_{S_1}$.
	The largest values of $R$\ are concentrated
	mostly in the region of quite large masses $m^{}_{S_1}$,
	where LEP constraints are not very significant.
	The upper bound for $m^{}_{S_1}$ is also clear seen. 	
		
\section{Conclusion}  \label{sec:conclusion}
	
	In the paper we address the question
	whether the Next-to-Minimal Supersymmetric Standard Model
	can be attractive from the point of view of the direct detection
	of neutralinos provided the neutralino is  the stable LSP.

	To answer the question we derived
	the effective low energy neutralino-quark Lagrangian, 
	which takes into account the contributions of extra 
	scalar Higgs boson and extra neutralino.
	On this basis we calculated the total 
	direct-dark-matter-detection event rate in $^{73}$Ge as a  
	representative isotope which is interesting 
	for construction of a realistic dark matter detector.	
	We analyzed the NMSSM taking into account
	the available accelerator and cosmological constraints
	by means of random scan of the NMSSM parameter space 
	at the Fermi scale.
	We demonstrated that the cosmological constraint does not 
	rule out domains in the parameter space which correspond to 
	quite sizable event rate in a germanium detector.

	Due to relaxation of the gaugino unification condition, 
	contrary to previous consideration
\cite{AStephan}
	we found domains in the parameter
	space where lightest neutralinos have quite small masses
	(about 3~GeV), acceptable relic
	abundance and sufficiently large expected event rate
	for direct detection with a $^{73}$Ge-detector.

	Therefore the NMSSM looks not worse then the MSSM from the
	point of view of direct dark matter detection.
	The question arises:
	Is it possible to distinguish MSSM and NMSSM 
	by means of direct dark matter detection of LSP? 
	It is a problem to be solved in future.
	The question can disappear by itself
	if negative search for light Higgs with LHC 
	rules out the MSSM.
	As already mentioned in the introduction 	
	the NMSSM can bypass the most crucial 
	constraint for the MSSM 
	with the upper bound 
	for the light Higgs boson 
(\ref{tlb}).
	Therefore the NMSSM might remain a viable theoretical 
	background for direct dark matter search for relic
	neutralinos in the post-MSSM epoch.

\bigskip

	We thank S.G. Kovalenko for helpful discussions.
	The investigation was supported in part (V.A.B.)
	by Grant GNTP 215 from Russian Ministry of Science
	and by joint Grant 96-02-00082 from Russian Foundation 
	for Basic Research and Deutsche Forschungsgemeinschaft.


\vfill 

\noindent	
{\bf Figure captions}

\bigskip 

\noindent	
{\bf Fig. 1:} 
	Total event rate $R$ for
	$^{73}$Ge, the LSP gaugino 
	fraction ($\CN^2_{11}+\CN^2_{12}$), 
	singlino fraction ($\CN^2_{15}$),
	and the
	relic abundance parameter $\Omega_\chi h^2_0$ 
	versus the LSP
	mass (from up to down).
	The left (right) panel presents results 
	obtained without (with) taking into account
	the cosmological relic density constraint.

\bigskip

\noindent	
{\bf Fig. 2:} 
	Total event rate $R$\ for $^{73}$Ge as function of
	the mass of the lightest scalar Higgs boson $m^{}_{S_1}$.

\vfill 

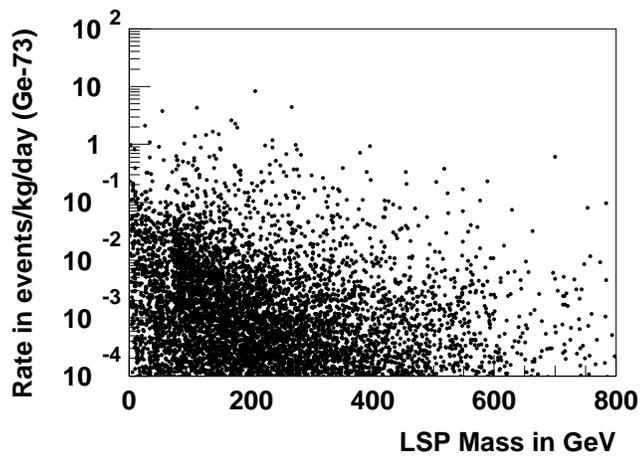
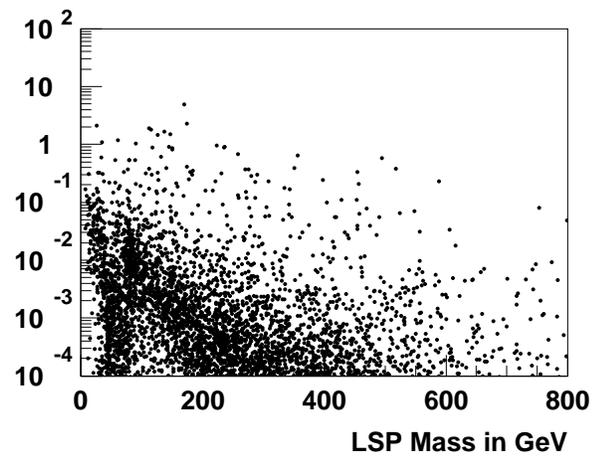
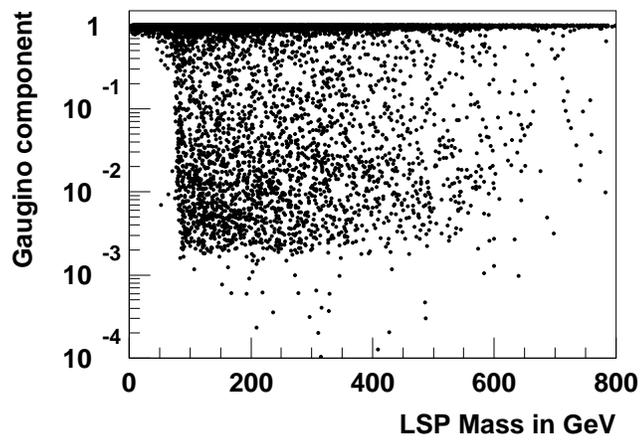
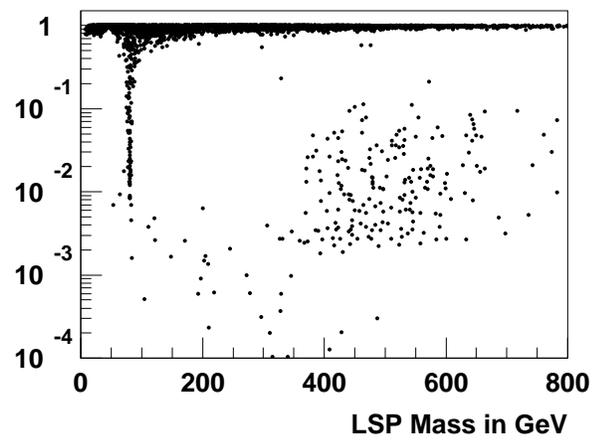
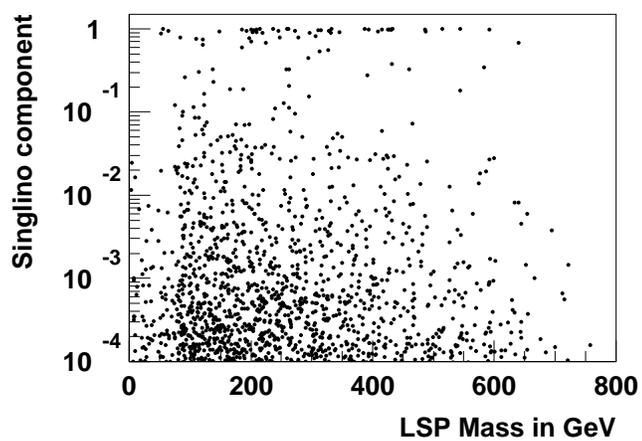
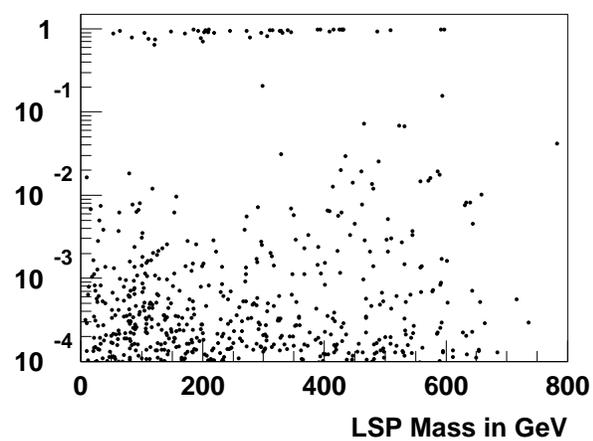
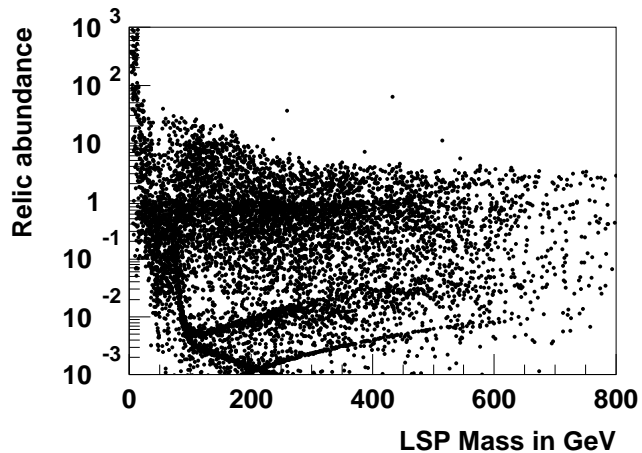
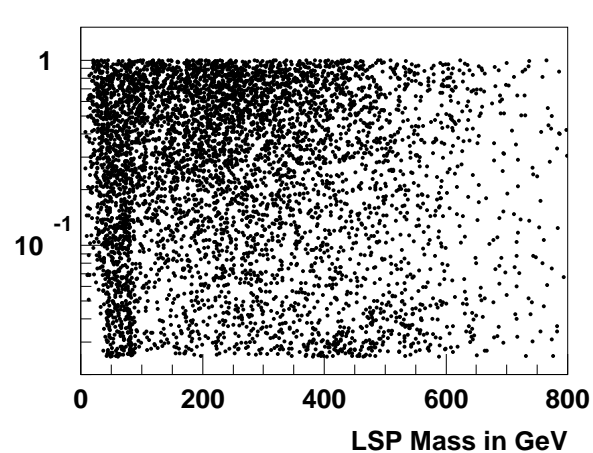
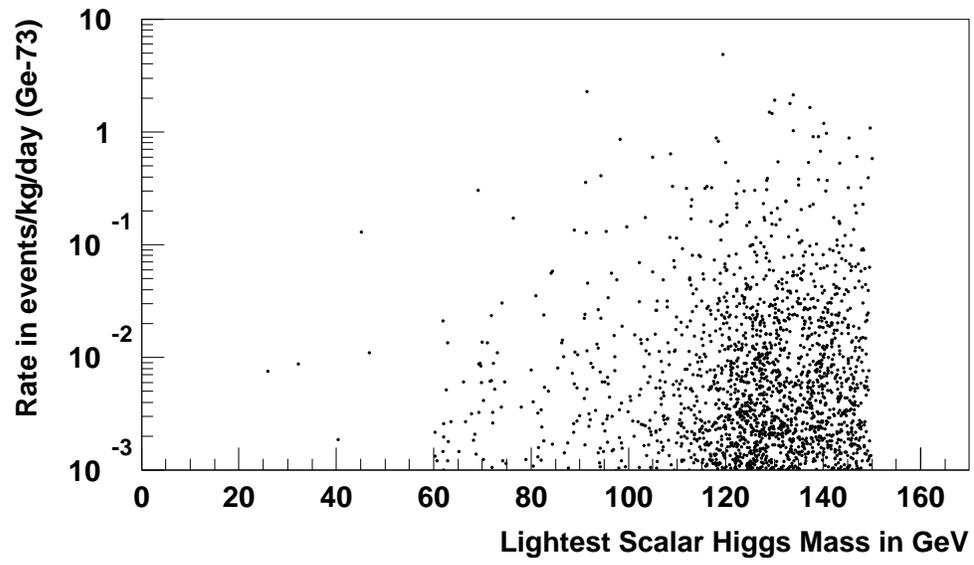
\end{document}